\newcommand{\eq}{\equiv}
\newcommand{\ze}{\emptyset}
\newcommand{\bumps}{B}
\newcommand{\fo}{f}
\newcommand{\rk}{\rho}
\newcommand{\ol}[1]{\overline{#1}}
\newcommand{\I}{Isabelle}
\newcommand{\myTitle}{$O_2$ is a multiple context-free grammar: an implementation-, formalisation-friendly proof}

\newcommand{\N}{\mathbb{N}}
\newcommand{\sdiff}{\backslash}

\newcommand{\reversed}[1]{\underline{#1}}
\newcommand{\rev}{\reversed}
\newcommand{\factor}[3]{ {\left \langle {#1} \right \rangle }_{#2}^{#3}}
\newcommand{\lf}{\preceq}

\documentclass
[
]
{llncs}

\usepackage{abraces}
\usepackage{cancel}
\usepackage{cite}
\usepackage{amsmath,amssymb,amsfonts}
\usepackage[inline]{enumitem}
\usepackage{graphicx}
\usepackage[]{hyperref}
\hypersetup{%
pdftitle={\myTitle{}}%
, pdfstartview=FitBH
, pdffitwindow=true
, pdfkeywords={models of computation,
formal languages,
computational algebra,
proof theory,
intelligent mathematics,
AI-aided mathematical discovery,
}
, pdflang=en-GB
, pdfpagemode=FullScreen
, pdfstartview= 100 100 10000
, pdfsubject={intelligent mathematics}
, pdfwindowui=true
, colorlinks=true
, linkcolor=blue
, urlcolor=blue
, citecolor=green
, breaklinks = true
, menucolor=brown
}
\usepackage[utf8x]{inputenc}
\usepackage[]{listings}
\usepackage[
]{microtype}
\usepackage{qtree}
\usepackage{tabulary}
\usepackage{times}
\usepackage{textcomp}
\usepackage{varwidth}
\usepackage{xcolor}

\title{\myTitle{}%
\author{
\href{http://orcid.org/0000-0002-4529-5442}
{Marco~B.~Caminati}\orcidID{0000-0002-4529-5442}
}
\institute{%
School~of~Computing~and~Communications
\\
Lancaster~University~in~Leipzig
\\
Nikolaistrasse 10
\\
04109 Leipzig
\\
Germany
\\
\email{m.caminati@lancaster.ac.uk}
}
}

\DeclareMathOperator{\argmin}{argmin}
\DeclareMathOperator{\length}{length}

\begin{document}
\maketitle{}
\begin{abstract}
Classifying formal languages according to the expressiveness of grammars able to generate them is a fundamental problem in computational linguistics and, therefore, in the theory of computation.
Furthermore, such kind of analysis can give insight into the classification of abstract algebraic structure such as groups, for example through the correspondence given by the word problem.
While many such classification problems remain open, others have been settled.
Recently, it was proved that $n$-balanced languages (i.e., whose strings contain the same occurrences of letters $a_i$ and $A_i$ with $1\leq i \leq n$) can be generated by multiple context-free grammars (MCFGs), which are one of the several slight extensions of context free grammars added to the classical Chomsky hierarchy to make the mentioned classification more precise.
This paper analyses the existing proofs from the computational and the proof-theoretical point of views, systematically studying whether each proof can lead to a verified (i.e., checked by a proof assistant) algorithm parsing balanced languages via MCFGs.
We conclude that none of the existing proofs is realistically suitable against this practical goal, and proceed to provide a radically new, elementary, extremely short proof for the crucial case $n \leq 2$.
A comparative analysis with respect to the existing proofs is finally performed to justify why the proposed proof is a substantial step towards concretely obtaining a verified parsing algorithm for $O_2$.
\end{abstract}

\pagebreak{}

\section{Introduction}
\label{RefSectIntro}
The classical connection between formal grammars, formal languages and abstract machines has been extensively and fruitfully explored. This led to an enhanced understanding of the capabilities and limitations of given models of languages and computability.
Firstly, by establishing links between the traditional models (e.g., regular grammars and finite state machines); subsequently, by introducing new models, usually for domain-specific reasons (e.g., computational linguists introducing new notions of formal grammars to better model natural languages) and studying how these new models translate.
This already fertile interplay has been made even more relevant, in later years, by noting how staple problems in computational abstract algebra (for example, combinatorial group theory, which in turn is a useful tool in geometry and topology) can be rephrased and solved using the definitions, ideas and methods of formal languages theory and computability theory~\cite{gilman2005formal}.

Therefore, the typical computational linguistics problem of establishing how powerful a formal grammar must be in order to generate and parse a given family of languages becomes relevant also in computability theory and computational abstract algebra.
One widely studied example is given by the family of $n$-balanced languages, containing exactly the words with equal pairwise occurrences of $n$ pairs of letters:
\begin{align}
\label{RefEqDefO}
O_n := \left\{  w \in \left(  \bigcup_{i=1}^{n} \left\{ a_i, \ol{a}_i \right\} \right)^*. \ 
 \ \forall i \leq n. \ 
\left| w \right|_{a_i} = \left| w \right|_{\ol{a}_i} \right\}.
\end{align}
Above, $|v|_{x}$ denotes the number of occurrences of the letter $x$ in a word $v$, and the alphabet $\Sigma_n$ is made up of the mutually distinct letters $a_i$'s and $\ol{a}_i$'s.
For a given $i$, we will say that $\ol{a_i}$ is the \emph{conjugate} letter of $a_i$ and vice versa.
The general notation $\ol x$ will indicate the conjugate letter of any given $x \in \Sigma_n$.
Another way to characterise a balanced language is via the introduction of the ancillary concept of \emph{balance}: given a string $w$ of $\Sigma_n$, its balance $\mu \left( w \right)$ is the $n$-tuple of integers 
$ \left( \left| w \right|_{a_1} - \left| w \right|_{\ol {a_1}}, \ldots ,
\left| w \right|_{a_n} - \left| w \right|_{\ol {a_n}}
\right)$, making a string balanced if and only if its balance has no non-zero entries.

From the point of view of computational linguistics, it is interesting to locate the $O_n$ family of languages within the known hierarchy of grammars proposed to model, e.g., natural languages.
From the point of view of computational group theory, information about formal languages can be used to explore corresponding groups~\cite{gilman2005formal}.
This led to a series of recent papers by several authors culminating with a result establishing that $O_n$ can be generated by a $n$-multiple context-free grammar ($n$-MCFG, see below for a definition) for any $n$.
All the existing proofs develop and improve on the original argument of Salvati~\cite{salvati2015mix}, being all essentially geometrical in nature, in stark contrast with the one given here.
While this geometric idea turned out to be powerful, it also has a couple of disadvantages: first, all its variations ultimately rely on key results which offer little indications as to which rules are used at every parsing step; secondly, as we will see, they would all likely turn out to be extremely hard to formalise in a proof assistant to check their validity.
This paper focuses on the computational and proof-theoretical aspects of that result, proposing a novel proof which has the advantage of being constructive and based on elementary ideas which are, in comparison, much more viable to a formalisation, and is the first step in a project aiming at delivering formally verified code which can parse $O_2$ words.
The price to pay for these upsides is that the new proof is limited to the bidimensional $n=2$ case which, however, is a  crucial case, having originated the series of papers mentioned above, and having significant interest on its own, both from a formal languages and from a computational group theory point of view given the relationship between $O_2$ and the open problems related to the word problem for $Z^2$~\cite{gilman2018groups}.
Section~\ref{RefSectMcf} introduces the problem and the needed definitions.
Section~\ref{RefSectProof} gives the basic idea behind the new proof and the remarkably short proof itself.
In Section~\ref{RefSectOldProofs}, we will systematically examine the existing proofs and their drawbacks in terms of implementability and formalisability.
Section~\ref{RefSectConclusion} concludes.

\begin{note}[Notation]
\label{RefNotation0}
In what follows, we will use the terms ``tuple'', ``list'' interchangeably; the terms ``string'' and ``word'' will refer to the same notion as ``tuple'' and ``list'', but will usually be adopted when working within the context of formal languages.
$\N$ will denote the set of natural numbers including $0$.
$|p|$ or $\length \left( p \right)$ is the length of the list $p$.
Sometimes we will regard a non-empty list as a function over an initial segment of $N$ (indices), thereby writing $p \left( i-1 \right)$ for the $i$-th entry (or the entry having \emph{index} $i$) of $p$, so that $p = p \left( 0 \right) p\left( 1 \right)\dots p\left( \left| p \right|-1 \right)$.
The notation $ \rev {p}$ will indicate the reverse of a given string $p$: i.e., $\rev {p} = p\left( \left| p \right|-1 \right) \dots p\left( 0 \right)$, and $\rev p \left( 0 \right)$ is the last letter of $p$ when $p$ is non-empty.
Finally, $\epsilon$ will denote the empty list.
\end{note}

\section{The context}
\label{RefSectMcf}
By the eighties, a general consensus formed around the idea that filling the gap between context-free grammars (CFGs) and context-sensitive grammars (CSGs) in the classical Chomsky hierarchy could have helped to more adequately model natural languages.
Under the loose term of \emph{mildly context-sensitive} grammars, several models were gradually conceived to add some expressiveness to context free grammars, to add some of the power of context-sensitive grammars in a somehow controlled way~\cite{kallmeyer2010parsing}.
As elaborated in Section~\ref{RefSectIntro}, this created new interesting problems which have been studied across disciplines, with one main problem among these being the generability of particular languages given a particular mildly context-sensitive grammar.
In Section~\ref{RefSectIntro}, we also anticipated which one of these (now solved) problems we are focusing on from a computability and proof-theoretical point of view: that of the generability of $O_n$ languages using the mildly context-sensitive grammar $n$-MCFGs.
With the easy definition of $O_n$ already given, we now focus on introducing MCFGs.

One way of looking at non-terminals in CFGs is as boolean functions of one string of terminals.
%
%
%
In 
a
parse tree, each non-terminal is the root of a subtree whose leaves form a substring of the final phrase. 
This can happen only for well-formed substrings (e.g., a valid noun phrase), and not for others; therefore, we can view at a given non-terminal as a function returning true exactly when this happens.

We also note that we can adopt this view of non-terminals as functions because the corresponding truth value does not depend on what is outside the given substring (this is given by context-freeness). 
Moreover, each substring matching a given non-terminal as in the example just given must be a contiguous substring of the final sentence%
.
This latter phenomenon corresponds to the fact that, in the view above of non-terminals as functions, these functions only take one argument~\cite{clark1985introduction}, and imposes substantial limitations on what a CFG can do.
To work around this limitedness, the main idea behind MCFGs is to allow such functions to take a finite number $n$ of arguments; then, $n$ is called the \emph{fanout}%
\footnote{Some sources adopt the term ``dimension'' in lieu of ``fanout''.}
of the corresponding non-terminal.
This strictly increases the expressiveness of the grammar, but without yielding the same expressiveness as CSGs~\cite{kallmeyer2010parsing}.
In particular, MCFGs allow to keep track of phrases with ``gaps'' in them~\cite{clark1985introduction} when parsing sentences.
Now, on top of this main idea, several details are to be sorted out: for example, how do non-terminals interact with each other, now that they all can have several arguments? 
And how does this issue relate to the notion of transitively applying rules to obtain derivations that we had in other grammars?
The following formal definition tackles these details.
As with, say, CFGs, it involves terminals, non-terminals, rules and a starting symbol.
Besides the discussed difference in the nature of non-terminals, the other main, consequent difference is in the structure of the rules.

\begin{definition}[MCFGs]
\label{RefDefMcfg}
A \emph{multiple context-free grammar} (MCFG) is given by four mutually disjoint, non-empty, finite sets $\Sigma$, $N$, $X$, $R$, by one distinct element $S \in N$, and by two maps $\fo$ and $\rk$ associating a natural number to, respectively, each \emph{non-terminal} of $N$ and to each \emph{rule} of $R$. 
The round brackets ``$\left(  \right)$'' arrow ``$\to$'' colon ``:'' and comma ``,'' symbols are not included in any of the sets above, as are neither $\fo$ nor $\rk$.
Each rule $r$ of $R$ has the form
\begin{small}
\begin{align}
\label{RefEqMcfg}
r: A \left( s_1, \ldots, s_{\fo\left( A \right)} \right) \to 
B_1 \left( x_{1,1}, \ldots, x_{1, \fo\left( B_1 \right)} \right),
\ldots,
B_{\rk\left( r \right)} \left( x_{\rk\left( r \right),1}, \ldots, 
x_{\rk\left( r \right),\fo\left(  B_{\rk\left( r \right)} \right)}
\right)
,
\end{align}
\end{small}
where $A$ and the $B_i$'s denote generic elements of $N$, and the following constraints hold:
\begin{enumerate*}
\item
all the $x_{i,j} \in X$ occurring in each rule are pairwise distinct; the set they constitute is denoted by $X_r$;
\item
$s_1 \ldots s_{\fo\left( A \right)} \in \left( X_r \cup \Sigma \right)^*$;
\item
each element of $X_r$ occurs exactly once in
$s_1 \ldots s_{\fo\left( A \right)}$;
\item
$\fo \left( S \right) = 1.$
\end{enumerate*}
$\fo$ is called the \emph{fanout} function and $\rk$ the \emph{rank} function.
$\underset{N}{\max}\ \fo$ is the fanout of the MCFG, and for any 
$n \ge \underset{N}{\max}\ \fo$, the grammar is said to be an $n$-MCFG.
$\Sigma$ is called the \emph{alphabet} of the MCFG, $X$ its set of \emph{variables}, and $S$ its \emph{start symbol}.
\end{definition}

Equation~\eqref{RefEqMcfg} is the key ingredient of Definition~\ref{RefDefMcfg}: it allows to interleave into the arguments $s_1, \ldots, s_{\fo\left( A \right)}$ the substrings held by the various $x$'s in the right-hand side, obtained by previous generative steps by using other non-terminals $B_1, \ldots, B_{\rk\left( r \right)}$.
This implies that, as usual, the language is generated by finitely iterating rule applications until non-terminals (except $S$) are gone.
This is made precise by the following definition.

\begin{definition}[Instance, sentential form, derivability]
\label{RefDefDer}
Given a MCFG $ G $ as the tuple $ \left( \Sigma, N , X, R, S, \fo, \rk \right)$, an \emph{instance} of any rule $r\in R$ is the string obtained by replacing both occurrences of each variable in the rule with one string of $ \Sigma^* $. 
Recursively, a \emph{sentential form} for $G$ is either the empty string or the left-hand side of an instance of some $r\in R$ such that all the comma-separated elements on its right-hand side are sentential forms.
A string $w$ of $\Sigma^*$ is \emph{derivable} (or a \emph{sentence}) in $G$ if $S\left( w \right)$ is a sentential form. 
The subset of $\Sigma^*$ constituted by all derivable strings in $G$ is called the language generated by $G$, indicated with $L\left( G \right)$, and styled a multiple context-free language (MCFL).
\end{definition}

We note that, while for CFGs it is customary to generate strings by applying rules left to right, with MCFGs, as from Definition~\ref{RefDefDer}, string generation is customarily obtained by applying rules right to left.
Consequently, to do the parsing of a string (as opposed to generating), one applies MCFG rules from left to right.
Additionally, we note that the recursive definition of a sentential form provided by Definition~\ref{RefDefDer} implies that the first step in generating strings must always begin from rules with an empty right hand side (base case).

We now introduce a specific MCFG, called $G_2$, for two reasons: first, it will act as a simple example to clarify the definitions above; secondly, we will use $G_2$ to generate the language $O_2$, thereby giving the proof central to this paper.
This means that the alphabet $\Sigma$ of $G_2$ coincides with the alphabet of the language $O_2$, which is obtained by substituting $n$ with $2$ in~\eqref{RefEqDefO}.
For notational convenience, however, we will denote its letters with $a, \ol a, b, \ol b$ in lieu of $a_1, \ol{a_1}, a_2, \ol{a_2}$, so that $O_2$ becomes
\begin{align}
O_2 := \left\{  w \in \left\{ a, \ol a, b, \ol b \right\}^*. \ \left| w \right|_{a} = \left| w \right|_{\ol{a}}, \ 
\left| w \right|_{b} = \left| w \right|_{\ol{b}}
\right\}.
\end{align}

\begin{definition}[$G_2$]
\label{RefDefG2}
Let $G_2$ be the MCFG with $\Sigma = \left\{ a, \ol a, b, \ol b \right\}$, $N=\left\{ S, I \right\}$,
$X=\left\{ v, w, x, y \right\}$, $R$ containing the following rules

\begin{tabular*}{\linewidth}{lcr}
\begin{varwidth}{.2\linewidth}
{\begin{align*}        
r_0		: I\left( \epsilon, \epsilon \right) & \to \\
r_a 		: I\left( a, \ol a \right) &\to\\ 
r_{\ol a} 	: I\left( \ol a, a \right) &\to\\
r_b 		: I\left( b, \ol b \right) &\to\\
r_{\ol b} 	: I\left( \ol b, b \right) &\to\\
\end{align*}
}
\end{varwidth}
&
\hphantom{sdfasfdasadfadfss}
&
\begin{varwidth}{0.5\linewidth}
{\begin{align*}
r_l 		: I\left( vxw, y \right) &\to I\left(v, w  \right), I\left(x, y  \right)\\
r_r 		: I\left( v, xwy \right) &\to I\left(v, w  \right), I\left(x, y  \right)\\
r_n 		: I\left( vx, yw \right) &\to I\left(v, w  \right), I\left(x, y  \right)\\
r_s 		: I\left( vx, wy \right) &\to I\left(v, w  \right), I\left(x, y  \right)\\
r_z 		: S\left( vw \right) & \to I\left( v, w \right)\\
\end{align*}
}
\end{varwidth}
\end{tabular*}
\vspace{-1.2cm}

with $\fo$ and $\rk$ as easily gatherable from the ruleset above.
\end{definition}
We want to prove is that \emph{any} string in $O_2$ admits a derivation in $G_2$, or, equivalently, that $ O_2 \subseteq L\left( G_2 \right)$.
This will prove the main result that $O_2 = L\left( G_2 \right)$ since the other set inclusion is obvious.
We start by noticing that, 
in general, 
there is not a unique derivation; in fact, we will prove a stronger result: for any pair of non-empty strings $\left( v, w \right)$ such that $vw \in O_2$, $I(v, w)$ is a sentential form.
This means that \emph{however} we choose to split \emph{any} non-empty string in $O_2$ into two proper factors, we can find a derivation where the second topmost node of the tree leads to the chosen split. 
While this result is stronger than what we need, it is liable to a straightforward structural recursion argument: it is easy to check that we will just need to prove that any $I \left( v,w \right)$ is the left hand of an instance of some rule of $G_2$, as soon as $vw \in O_2$ and $\left| v \right|, \left| w \right|>0$.
This is immediate for strings with length not exceeding two, so that we can restrict our attention to longer ones.
We also note that the special rule $S$ will only be used at the very final step of the string generation, with only $I$ playing a non-trivial role.
As a consequence, we will just drop the symbol $I$ to declutter the notations in our proofs.
With this simplified notation, the result we need to prove is the following main theorem.

\begin{theorem}
\label{RefLmMain}
Any proper, binary factorisation of a string of $O_2$ of length bigger than $2$ admits a balanced decomposition.
In other words, the set
$
\{ \left( z_0, z_1 \right). \ z_0 z_1 \in O_2, \ \left| z_0 z_1 \right| > 2 , \ \min\left\{ \left| z_0 \right|, \left| z_1 \right| \right\} >0, 
\left( z_0, z_1 \right)\text{ admits no balanced decomposition} \}
$
is empty.
\end{theorem}

In Theorem~\ref{RefLmMain}, the notion of balanced decomposibility reproduces that of derivability, but without any reference to a particular rule, in accordance to our decluttered notation introduced above:

\begin{definition}[Factorisation, decomposition]
\label{RefDefDecomp}
A \emph{factorisation} of a string $s$ is a finite list of strings $\left( s_0, \ldots, s_{n-1} \right)$ such that $s=s_0 \ldots s_{n-1}$; 
each of its entries is called a \emph{factor} of $s$; a left factor if it is the first entry; a right factor if it is the last; a \emph{proper} factor if it is not empty%
; \emph{ultra-proper} if it is proper and distinct from $s$; the factorisation is proper if each of its entries is proper.
\\
Given a finite, non-empty list of non-empty strings $\left( s_0, \ldots, s_{n-1}  \right)$, 
choose a proper factorisation for each $s_i$, and put all the resulting factors into a list $\omega$, in some arbitrary order. 
If $\omega$ can be partitioned
into two lists $\left( p_0, \ldots, p_{l-1}  \right)$ and 
$\left( q_0, \ldots, q_{m-1}  \right) $ with $0< l, m \leq n$ and $\left| p_0 \ldots p_{l-1} \right|, \left| q_0 \ldots q_{m-1} \right| < \left| s_0 \ldots s_{n-1} \right|$, we will say that $\left( \left( p_0, \ldots, p_{l-1} \right), \left( q_0, \ldots, q_{m-1} \right) \right)$ is a \emph{decomposition} of $\left( s_0, \ldots, s_{n-1}  \right)$.
In this case, we will call each list  $ \left( p_0, \ldots, p_{l-1} \right)$ and $ \left( q_0, \ldots, q_{m-1} \right)$ a \emph{component} of the decomposition.
\\
A factorisation of a balanced word will also be called balanced.
Finally, a decomposition the components of which are all balanced will be called balanced or a \emph{bal-decomposition}.
\end{definition}

Note that Definition~\ref{RefDefDecomp} imposes that, to be decomposed, a list of strings must be a proper factorisation of some string. 
Additionally, it imposes that any of its components is also a proper factorisation of some string, with the additional condition of its length being bigger than $0$ and at most the length of the original list.
Finally, observe that the operation of factorisation applies to a string and returns a list of strings, while that of decomposition applies to a list of strings and return a list of lists of strings.

\begin{note}[Notation]
\label{RefNotation1}
Occasionally, it will be convenient to denote a factor of a given non-empty string $q$ using the correspondent set of indices, by writing $\factor{q}{i}{j}$ (or, alternatively, $\factor{q}{\left[ i, j \right]}{}$) to mean the factor $\Pi_{k=i}^{\min \left\{ j, \left| q \right| -1 \right\}} q \left( k \right)$ for $i, j \in \N$.
Note that this yields the empty string for some values of $i, j$, for example when $j<i$.
We will just write $\factor{q}{}{j}$ in lieu of $\factor{q}{0}{j}$, and 
$\factor{q}{i}{}$ instead of $\factor{q}{i}{\left| q \right|}$.
Additionally, we will use the notation $p - X$, where $X\subseteq \N$, to indicate the string obtained by removing all letters having indices in $X$: for example, 
$a b \ol b b \ol a - \left\{ 2, 4 \right\} = 
a b b $.
Finally, the notation $p \lf q$ will mean that $p$ is a left factor of $q$.
\end{note}
The following propositions and corollaries (proved in Appendix~\ref{RefSectApp}) establish the sufficiency of Theorem~\ref{RefLmMain} to obtain the main result that $L \left( G_2 \right) = O_2$ making thus $O_2$ a MCFL. 
Therefore, the rest of the paper, starting with next section, will be devoted to proving Theorem~\ref{RefLmMain}.

\begin{proposition}
\label{RefPropMain}
Given a factorisation $\left( p, q \right)$ of a string of $O_2$, $I\left( p, q \right)$ is a sentential form for the grammar $G_2$ of Definition~\ref{RefDefG2}.
\end{proposition}

\begin{corollary}[of~Proposition~\ref{RefPropMain}]
\label{RefLmG2O2}
$ L \left( G_2 \right) = O_2$.
\end{corollary}

\section{The new proof}
\label{RefSectProof}
\begin{definition}
\label{RefDefShort}
Consider the equivalence relation $\eq_n$ (or just $\eq$ when clear) induced over $\Sigma_n$ by the balance map $\mu$ introduced in Section~\ref{RefSectMcf}: that is, $\Sigma_n$ is partitioned into classes of strings whereby two strings are in the same class if and only if they have the same balance.
Each class will have a subset of strings of minimal length, which we call \emph{short}.
In other words, a short string is one not having two letters from the same pair occurring: if an $a_i$ occurs in it, then $\ol {a_i}$ does not occur, and vice versa.
A list made up entirely of short strings will also be called short.
\end{definition}

Now, going back to the case $n=2$, we start from the restriction of Theorem~\ref{RefLmMain}
to the short case (recall from Note~\ref{RefNotation0} that the notation $\rev{x}$ denotes the reversion of the list $x$, and, in particular, $\rev{x} \left( 0 \right)$ is the last entry of $x$ if the latter is non-empty):

\begin{lemma}
\label{RefLmShort1}
Let $\left( x, y \right)$ be a proper, short factorisation of a string of $O_2$ of length $>2$.
Then, either $\ol{x\left( 0 \right)} = \rev{y} \left( 0 \right)$, or $
\rev{x} \left( 0 \right) = \ol{y \left( 0 \right)}
$, or there are $p$ and $q$ ultra-proper left factors of $x$ and $y$, respectively, and such that $pq \in O_2$.
\end{lemma}
\begin{proof}
Assume the theorem does not hold, and consider among the counterexamples a particular $\left( x,y \right)$ such that $\left| xy \right|$ is minimal.
By shortness and without loss of generality, we can then assume $x=a \alpha x'$ and $y=\ol b^{1+l} \ol a y' \ol b$ for some $l\in \N$, with $\alpha \in \left\{ a,b \right\}$, $x' \in \left\{ a, b \right\}^*$ and $y' \in \left\{ \ol a, \ol b \right\}^*$.
Now if $\alpha$ is $a$, the thesis holds for $\left( \alpha x', \ol b^{1+l} y' \ol b \right)$ by minimality, implying that there must be ultra-proper left factors $p$ and $q$ of $a x'$ and $\ol b^{1+l} y' \ol b$ respectively such that $pq$ is balanced.
But then it would be $\left| q \right| > 1+l $ in order to compensate the first letter ($a$) of $p$, so that we can easily obtain a decomposition of $\left( x,y \right)$, contradicting our initial assumption.

Therefore, it must be $\alpha = b$ and $l=m+1$ for some $m \in \N$.
Then, there must be two ultra-proper left factors $p$ and $q$ of $a x'$ and of $\ol b^{m+1} \ol a y' \ol b$ respectively such that $pq$ is balanced.
Again, then $\left| q \right| > m+1$ and $\left| p \right| > 1$, so that we can easily extend $p$ and $q$ to ultra-proper left factors of $x$ and $y$, respectively, contradicting our initial assumption.

\end{proof}

\begin{corollary}
\label{RefLmShort2}
Any proper, short, binary factorisation of a string of $O_2$ of length bigger than $2$ admits a balanced decomposition.
\end{corollary}
\begin{proof}
Lemma~\ref{RefLmShort1} asserts that it is possible to find such a decomposition of a very special form (equivalent to using either $r_s$ or a restricted form of $r_n$).
Hence, the corollary follows immediately.
\end{proof}

Now that we settled the problem for the short case with Corollary~\ref{RefLmShort2}, we can use it as a base case for the, general, non-short case.
The idea is to reduce the latter to the former by eliminating pairs of conjugate letters.
The following definitions gives us a way to do this in the most gradual way.

\begin{definition}
\label{RefDefBump}
A minimal non-short factor of a string $p \in \Sigma_n^*$ will be called a bump of $p$, and will be indicated through its indices.
More formally, $\left[ i, j \right]$ is an $x$-bump (or just bump) for $p$ if $\factor{p}{i}{j}$ is not short, all its ultra-proper factors are short, and $p \left( i \right) = x$.
Note that this implies $p\left( j \right)= \ol x$, and that 
$ p \left( k \right) \notin \left\{ x, \ol x \right\} \forall k \in \left] i, j \right[$.
$x$ is called the \emph{direction} of the bump.
$\bumps^x \left( p \right)$ is the set of all $x$-bumps of $p$, and we set 
$ \bumps^X \left( p \right) := \bigcup_{x \in X} \bumps^x \left( p \right)$,
$ \bumps \left( p \right) := \bigcup_{x \in \Sigma_n} \bumps^x \left( p \right)$.
If $\left[ i, j \right]$ is a bump for $p$, the string $ \factor p i j $
will also be called a bump for $p$.
\end{definition}

\begin{note}
$p$ is short if and only if $\bumps \left( p \right) = \ze$.
Also note that, in the case $n=2$, any bump has the form $\alpha \beta^m \ol \alpha$ for some $m \in \N$, with $\alpha$ and $\beta$ non-conjugated letters of $\Sigma_2$.
\end{note}

\subsection{The non-short case: intuition}
\label{RefSectIntui}
We will now suppose there is a balanced but not bal-decomposable factorisation of a string of $O_2$ longer than $2$ (that is, violating Theorem~\ref{RefLmMain}), so that we can pick one such factorisation having a concatenation of minimal length.
We first note that, by minimality, nowhere in this factorisation a bump of length $2$ (that is, of the form $\alpha \ol{\alpha}$) can occur: this is easy to see and will be later formally implied by Proposition~\ref{RefThmRepl}.
Secondly, at least one of the two component strings (let us say the first one) will have a bump by Corollary~\ref{RefLmShort2} and Definition~\ref{RefDefBump}, so that we can pick one of minimal length and cancel its two extrema, thus preserving balancedness of their concatenation.
By minimality, the ``canceled'' factorisation is now bal-decomposable, so that we have the situation depicted below (where the parentheses enclose each component of the factorisation):
\begin{align*}
 ( \aunderbrace[,2']{p_0 \cancel{a} b \dots b \hphantom{|}}[D]_{A} \dots b \cancel{\ol a} p_1 \aunderbrace['2,]{p_2}[D]_{B} ) ( p_3 ).
\end{align*}
\\
We have used under-brackets to mark one of the two components of a bal-decomposition of the ``canceled'' factorisation. 
This means that the portion marked by the brackets is balanced, as is the unmarked one.
Note that one bracket (here indicated with $A$) must straddle the canceled bump (otherwise we could reinstate the two canceled letters and still have a bal-decomposition), while the other ($B$) must not (otherwise we could again reinstate the two canceled letters and still have a bal-decomposition).
The other bracket ($B$) could also mark a left or right factor of $p_3$ depending on the particular bal-decomposition: this would not change the following reasoning.
We also assumed, without loss of generality, that the direction of the bump is $a$ (see Definition~\ref{RefDefBump}).

Now, the first letter of $p_2$ cannot be $a$, otherwise we could reinstate the canceled letters and obtain a bal-decomposition. 
Similarly, the last letter of $p_1$ cannot be $\ol a$.
There are two cases: the last letter of $p_1$ is $a$ or $b$ (the case $\ol b$ is similar to the latter).
In the first case, the first letter of $p_2$ must be in $\left\{ b, \ol b \right\}$, let us say it is $b$ (again, the case $\ol b$ is similar):

\begin{align*}
( \aunderbrace[,2']{p_0 \cancel{a} b \dots b \hphantom{|}}[D]_{A} \dots b \cancel{\ol a} p'_1 a \aunderbrace['2,]{b b \dots p'_2}[D]_{B} ) ( p_3 )
\end{align*}

In this configuration, we can begin ``sliding'' the upwards edge of $A$ and the upwards edge of $B$ rightwards one letter at a time while preserving balancedness. 
If the upwards edge of $A$ reaches the canceled $\cancel{\ol a}$, we can, again, reinstate the canceled letters.
Otherwise, after several rightwards sliding steps, we will eventually find ourselves (supposing $B$ has not shortened to length $0$) in a configuration like this:

\begin{align*}
( \aunderbrace[,2']{p_0 \cancel{a} b \dots b b b \dots b \hphantom{|}}[D]_{A} \dots b \cancel{\ol a} p'_1 a b b \dots b \aunderbrace['2,]{ \ol{a} p''_2}[D]_{B} ) ( p_3 ).
\end{align*}
Note that the leftmost letter under $B$ cannot be $b$ (otherwise we could have kept sliding), cannot be $a$ (otherwise we could reinstate the canceled letters and obtain a bal-decomposition), and cannot be $\ol b$ because, as previously observed, there cannot be bumps of length $2$: hence it must be $\ol a$, as depicted above.
But this means that the bump on the right in the figure above is of length strictly smaller than the canceled bump, which is impossible because we chose one of minimal length.

The case where the last letter of $p_1$ is $b$ entails a similar sliding reasoning, and is left to the reader.
Please note that this informal explanation leaves a number of corner cases unexplored which will be formally addressed in the forthcoming lemmas and propositions.
For instance, in our example, the reasoning becomes problematic when the canceled bump touches one or both extrema of the factor (that is, the round brackets in the figures above): this is addressed in the last proof of this paper.
Or, during the ``sliding'', the bracket $B$ could collapse to a length of $0$, which is addressed in Proposition~\ref{RefThmRepl}.
The discussion above was chiefly added to help the reader understanding the remaining formal results below.

\subsection{The non-short case: formal proofs}
We start from a result valid not only in the case $n=2$, but in general. 
It states that balanced-decomposability is preserved upon replacing factors of length smaller than $2$ with equivalent ($\eq_n$ as from Definition~\ref{RefDefShort}) ones.
In particular, this implies that a b-irreducible factorisation cannot contain bumps of length $2$ or $3$, a fact that we have already used in our informal ``sliding'' reasoning above.
It will be used repeatedly in the subsequent proofs.

\begin{proposition}
\label{RefThmRepl}
Assume $\left( s_0, \ldots, s_{n-1} \right)$ is balanced-decomposable in $\Sigma^*_n$, and that $s \eq \factor{s_0}{i}{i-1+\delta}$ with $\delta \in \left\{ 0, 1 \right\}$.
\\
Then $\left( \factor{s_0}{}{i - \delta} s \factor{s_0}{i+1}{}, \ldots, s_{n-1} \right)$ is also balanced-decomposable.
\end{proposition}
\begin{proof}
We can assume that the entry of index $i$ of $s_0$ is the entry of index $j$ of $p_0$, where 
$\left( p_0, \ldots, p_{l-1} \right)$ is a component of a balanced decomposition of 
$\left( s_0, \ldots, s_{n-1} \right)$.
Then $p_0 \eq p_0':= \factor{p_0}{}{j-\delta} s \factor{p_0}{j+1}{}$, so that
$\left( p_0', \ldots, p_{l-1} \right)$, $\left( q_0, \ldots, q_{m-1} \right)$ is a decomposition of
$\left( \factor{s_0}{}{i - \delta} s \factor{s_0}{i+1}{}, \ldots, s_{n-1} \right)$.
\end{proof}

The following lemma expresses formally the intuitive ``sliding'' argument of Section~\ref{RefSectIntui}.
More precisely, it states that it can only fail when the ``canceled'' bump is a left or right factor of a factor of a balanced word.

\begin{lemma}
\label{RefLmBump}
Let $s_0$, $s_1$ be non-empty strings of $\Sigma_2^*$.
Assume that 
\begin{enumerate}
\item
\label{RefHypMinBump}
$ \mathcal{B} := [i,1+i+k] \in \bumps^{\alpha}\left( s_0 \right) \cap 
\argmin_{\length} \left( \bumps^{\left\{ \alpha, \ol \alpha \right\}} \left( s_0 \right) \cup \bumps^{\left\{ \alpha, \ol \alpha \right\}} \left( s_1 \right) \right)$, 
\item
\label{RefHypBad}
$\left( s_0, s_1 \right)$ is balanced (recall that this implies $s_0 s_1 \in O_2$) but not bal-decomposable, 
\item
\label{RefHypMinBad}
for any $p_0$ and $p_1$ being substrings of $s_0$ and of $s_1$, respectively, such that $p_0 p_1 \in O_2$ and $\left| p_0 p_1 \right| < \left| s_0 s_1 \right|$, it holds that $\left( p_0, p_1 \right)$ is balanced-decomposable, and
\item
\label{RefHypNoBorder}
$k < \left| s_0 \right| -1$.
\end{enumerate}
Then $i=0$, $k \geq 2$, and ${\ol{s_0 \left( i+1 \right)}}^{j} \alpha \lf x$ for some $x \in \left\{ \rev s_0, s_1, \rev s_1 \right\}$ and $1 \leq j < k$. 
\end{lemma}

\begin{proof}
Note that Proposition~\ref{RefThmRepl}, together with hypotheses~\eqref{RefHypMinBad} and~\eqref{RefHypBad}, allows us to infer that 
\begin{align}
\label{RefEqBigBumps}
\forall \left[ i', i'+1+j' \right] \in \bumps\left( s_0 \right) \cup \bumps\left( s_1 \right). \ j' \geq 2, 
\end{align}
and, in particular, $k \geq 2;$ then, $s_0$ must be of the form 
$s_2 \alpha \beta^{1+l} \beta^{1+m} \ol \alpha s_3$, 
with $\beta := s_0 \left( i+1 \right) \notin \left\{ \alpha, \ol \alpha \right\}$ and $s_3 \neq \ze$ thanks to hypothesis~\eqref{RefHypNoBorder}.
Hypothesis~\eqref{RefHypBad} tells us that $\left| s_0 s_1 \right| > 4$ by a straightforward check.
Therefore, using hypothesis~\eqref{RefHypMinBad}, the set of $D$ of bal-decompositions of 
$\left( s_0' := s_0 -\left\{i, 1+i+k \right\}, s_1 \right)$ (where we used the $-$ notation introduced in Note~\ref{RefNotation1}) is not empty.
Additionally, any decomposition in that set will be made up of two components each of length $2$ since, if that were not the case, we could use Proposition~\ref{RefThmRepl} and hypothesis~\eqref{RefHypMinBad} to violate hypothesis~\eqref{RefHypBad}; to recapitulate:
\begin{align}
\label{RefEqDecomp}
D \neq \ze \text{ and } \forall x \in D. \ x = \left( \left( p_0, q_0 \right), \left( p_1, q_1 \right) \right) \text{ with } p_0, q_0, p_1, q_1 \neq \epsilon.
\end{align}
Furthermore, any bal-decomposition in $D$ must factor $s_0$ into at least two non-empty strings ``straddling'' the bump $\mathcal{B}$, otherwise, again, we would easily violate hypothesis~\eqref{RefHypBad}.

Let us then pick a $\left( \left( p_0, q_0 \right), \left( p_1, q_1 \right) \right) \in D$.
Without loss of generality, there are only two cases: 
\begin{enumerate*}
\item
$s_0 -\left\{i, 1+i+k \right\} = p_0 p_1 q_0, $ and 
\item
\label{RefCaseTwo}
$s_0 -\left\{i, 1+i+k \right\} = p_0 p_1$.
\end{enumerate*}
The proofs in the two cases are similar; let us do the second,%
\footnote{The reader is reminded that the first case was informally illustrated in Section~\ref{RefSectIntui}.}
so that (maybe after one reversal of $s_1$, which would pose no problem since reversion does not interfere with bal-decomposability) $s_1 = q_0 q_1$.
We can assume that our $q_0$ has minimal length among all the choices of $p_0, q_0, p_1, q_1$ satisfying case~\ref{RefCaseTwo} above.
Due to the above observation about straddling, $p_0 = s_2 \beta^{1+l}$ with $\left| s_2 \right|=i$.
By~\eqref{RefEqDecomp}, we can consider the rightmost letter of $q_0$ and the leftmost one of $q_1$, calling them $y$ and $z$, respectively.
By minimality of $\left| q_0 \right|$, and since $s_3$ is non-empty, it cannot be $y = \beta$. 
Moreover, $y \neq \alpha$ due to hypothesis~\ref{RefHypBad}.
We now show that $y \neq \ol \alpha$:
\begin{description}
\item[proof that $y \neq \ol \alpha$:]
if $y$ were $\ol \alpha$, then $z$ could not be $\alpha$ due to~\eqref{RefEqBigBumps}, and could not be $\ol \alpha$ due to hypothesis~\eqref{RefHypBad}.
\begin{description}
\item[case $z=\beta$:] then we could write $q_1$ as $\beta^{1+l'} q'_1$ with $l' < l$ due to hypothesis~\eqref{RefHypBad}, and $q'_1$ not having $\beta$ as its first letter.
Note that 
\begin{align}
\label{RefEqBeta}
\left( \left( s_2 \beta^{l-l'}, q_0 \beta^{1+l'} \right), \left( \beta^{2+m+l'} s_3, q'_1 \right) \right) \in D,
\end{align}
so that $q'_1$ cannot be empty due to~\eqref{RefEqDecomp}. 
Its first letter cannot be $\alpha$ due to $l'< l$ and hypothesis~\eqref{RefHypMinBad}, and it cannot be $\ol \beta$ due to~\eqref{RefEqBigBumps}.
It cannot be $\ol \alpha$ due to~\eqref{RefEqBeta} and hypothesis~\eqref{RefHypBad}. 
Contradiction.
\item[case $z= \ol \beta$: ]
then we could write $q_1$ as $\ol \beta^{1+m'} q'_1$ (with $m' < m$ due to hypotheses~\eqref{RefHypBad} and~\eqref{RefHypNoBorder}), and $q'_1$ not having $\ol \beta$ as its first letter.
Note that 
\begin{align}
\label{RefEqCoBeta}
\left( \left( s_2 \beta^{2+l+m'}, q_0 {\ol \beta}^{1+m'} \right), \left( \beta^{m-m'} s_3, q'_1\right) \right) \in D,
\end{align}
so that $q'_1 \neq \epsilon$ by~\eqref{RefEqDecomp}.
Its first letter cannot be $\alpha$ due to $m' < m$ and hyopthesis~\eqref{RefHypMinBad}, and cannot be $\beta$ due to~\eqref{RefEqBigBumps}.
It cannot be $\ol \alpha$ due to~\eqref{RefEqCoBeta} and hypothesis~\eqref{RefHypBad}.
Contradiction.
\end{description}
\end{description}
We conclude that $y=\ol \beta$ and hence $q_0= \ol \beta$ by minimality of $\left| q_0 \right|$, from which it follows that $l=0$ and $s_2=\epsilon$ by~\eqref{RefEqDecomp}: otherwise, we would have that
\\
$\left( \left( s_2 \beta^l, \beta^{2+m}  s_3 \right), \left( q_0', y q_1  \right) \right)$, 
where 
$q_0 = q_0' y$, is a bal-decomposition of $\left( s_0' , s_1 \right)$, against the minimality of $\left| q_0  \right|$.
It is now easy to check that $q_1$ must contain at least one letter different from $\ol \beta$, so that we can write $q_1 = {\ol \beta}^{h} \gamma q'_1$ for some $h\in \N$, $\gamma \in \left\{ \alpha, \ol \alpha, \beta \right\}$, and $q'_1 \in \Sigma_2^*$.
As usual, $\gamma \neq \beta$ by~\eqref{RefEqBigBumps}.
Moreover, $h < m+1 < k \Rightarrow h+1 < k$: otherwise, since $s_3 \neq \epsilon$, we could violate~\eqref{RefHypBad}.
Therefore, $ \left( \left( \beta^{1+h}, {\ol \beta}^{1+h} \right), \left( \beta^{m-h+1} s_3, \gamma q'_1 \right) \right) \in D$.
It follows that $\gamma$ cannot be $\ol \alpha$, otherwise we would obtain a bal-decomposition of $\left( s_0, s_1 \right)$; 
hence, $\gamma = \alpha$, terminating the proof.
\end{proof}

We are now in a position to perform our final proof.

\begin{proof}[of Theorem~\ref{RefLmMain}]
Assume that the set in the statement of Theorem~\ref{RefLmMain} is non-empty, and choose in it some $\left( q_0, q_1 \right)$ such that $\left| q_0 q_1 \right|$ is minimal.
$q_0$ and $q_1$ cannot both be short by virtue of Lemma~\ref{RefLmShort2}. 
Therefore, by Lemma~\ref{RefLmBump} and without loss of generality, we can assume that $a b^{2+m'} \ol a \lf q_0$ and that ${ \ol b}^{1+m} a \lf x$ for some $0 \leq m \leq m'$, $x \in \left\{ \rev{q_0}, q_1 \right\}$.
\begin{description}
\item[Case $x=\rev{q_0}$: ]
then there is a minimal $\beta$-bump either in $q_0$ or $q_1$, where $\beta \in \left\{ b, \ol b \right\}$.
It is easy to check it cannot be in $q_0$ by reapplying Lemma~\ref{RefLmBump}.
Therefore, it must be in $q_1$ and, again by Lemma~\ref{RefLmBump}, we can assume it is a left factor of $q_1$, so that $ \beta {\ol a}^{1+i} \ol \beta \lf q_1$, with $\beta \in \left\{ b, \ol b \right\}$: note that the second letter of $q_1$ cannot be $a$, otherwise Lemma~\ref{RefLmBump} would make $\left( q_0, q_1 \right)$ bal-decomposable.
But now, whichever value we choose for $\beta$, $\left( q_0, q_1 \right)$ would be bal-decomposable.
\item[Case $x=q_1$: ]
it is easy to check that, by Lemma~\ref{RefLmBump}, neither $q_0$ nor $q_1$ can have any $\beta$-bump, where $\beta \in \left\{ b, \ol b \right\}$. 
This implies that the last letter of $q_0$ is $a$.
We can pick the minimal $m$ making $\left( q_0, q_1 \right)$ bal-indecomposable.
Then, by calling $q_1'$ the string obtained from $q_1$ by swapping the $\left( 1+m  \right)$-th $\ol b$ and the contiguous $a$ in $q_1$, $\left( q_0, q_1' \right)$ will become bal-decomposable, and any bal-decomposition will split $q_1'$ immediately after the leftmost $a$ of $q_1'$.
Since $m \leq m'$ and the occurrences of $\ol b$ in $q_0$ and of $b$ in $q_1'$, respectively, are zero, the only possibility is that there is a right factor $q_2$ of $q_0$ such that $q_2 \ol b^{m}a$ is balanced. 
But this would imply $q_2$ containing an $a$ or $\ol a$-bump, and therefore at least $1+m' > m$ occurrences of $b$, which is clearly impossible, since $q_2$ contains no $\ol b$.
\end{description}
\end{proof}

\section{Comparison with the existing proofs}
\label{RefSectOldProofs}

In the peer-reviewed literature, there are four proofs of results relating to Theorem~\ref{RefLmMain}, with various degrees of generality.
Contrary to the present proof, they are all derivations or improvements of the original proof in~\cite{salvati2015mix}, from which we start our comparison.
The two main criteria guiding our comparison are: 
\begin{enumerate*}
\item
\label{RefItemFeas}
the feasibility of a concrete formalisation of the proof in a modern proof assistance language (e.g., \I{}/HOL) and 
\item
\label{RefItemConstr}
the performance of a corresponding parser obtained from the definitions involved in the formalisation.
\end{enumerate*}
To help assessing the first criterion, we will consider the requirements of each proof in terms of the mathematical definitions and results assumed as given, and the size in bytes of the textual proof.
The second aspect has some elements of arbitrariness and discretion, given the somewhat difficult task of deciding what to extract from a paper to consider as part of a proof and what not. 
However, although crude, this byte counting approach is a well-established method when quantitatively analysing mathematical proofs, for example to establish their de~Bruijn factor in the context of mechanised proving~\cite{asperti2010some,naumowicz2006example}.
In measuring the byte-size of paper proofs, the explanatory notes and the parts addressed to the reader, (as opposed to the proofs, statements, and definitions) were not considered. 
Technically, the size was measured by selecting the relevant text in the papers' pdf files.
For the proof contained in this paper, this gives a result of around $10$kB.

Regarding criterion~\eqref{RefItemConstr}, 
the mere existence of results, such as the one in the present paper, linking $O_2$ and some MCFG makes the membership problem for the given grammar trivial.
The parsing problem remains more complicated: since we know that any string in $O_2$ is derivable in $G_2$, we can brute-force our way through a given string to build a parse tree, or use MCFG-dedicated algorithms to generate a parse tree~\cite[Section~7]{kallmeyer2010parsing}.
However, these algorithm tend to be complicated to implement and hence hard to verify, with complexities typically exceeding $O \left( n^6 \right)$.
On the other hand, Lemma~\ref{RefLmMain} provides highly useful information to avoid brute forcing: it tells us that we can cancel a minimal bump and use the decomposition of the reduced string, instead, which gives the basis for a recursive parsing.
Thus, the parsing problem can be recursively reduced to the simple problem of finding a list of bumps for a given string and book-keeping it as long as they are canceled one by one.
We will now analyse whether the existing proofs provide any similar aspect.

\textbf{The original proof by Salvati.}
The proof in~\cite{salvati2015mix} is geometrical and involves non-elementary notions such as that of homotopy, winding numbers, fundamental groups, covering spaces.
Some facts, such as the unique path-lifting property and the homotopy-lifting property, are stated without proof.
The remaining proofs and definitions (excluding MCFG-related definitions and proofs not strictly needed for the main proof, such as Lemma~1) amount to about $83$kB.
For these reasons, the proof looks very hard to formalise.
The theorem on Jordan curves proved in Section~4 is highly non-constructive, for example resorting to Zorn's lemma to prove Lemma~13, thereby providing little insight from the parsing point of view.
\\
\textbf{The proof by Nederhof.}
In~\cite{nederhof2016short}, Nederhof provides a shorter proof than that in~\cite{salvati2015mix}.
This avoids the need of problematic (from a formalisation point of view) dependencies such as homotopy, fundamental groups, covering spaces.
However, many ideas are shared between the two proofs, including representing words as curves in the plane, using them to pass to the continuous geometrical view, and then finding a way to tame the extremely complicated curve's self-intersections which may arise.
This implies that similar difficulties would arise with respect to the formalisation.
For the same reason, building an efficient algorithm from this proof looks quite unfeasible.
Another issue is that the proof given is quite unstructured (no separation into lemmas, theorems, propositions, etc.) and, mostly important, quite informal, with abundance of appeals to geometrical intuition (e.g., ``it is clear that
$m$, $Q'_1$ and $Q_2'$ satisfying these requirements must exist'', or 
``The truncation consists in changing $sub_C (Q'_1 , Q'_2)$ to become seg$(Q'_1 , Q'_2 )$, as illustrated by Figure 7.''
).
While this kind of style is likely readily accepted and appreciated by a human reader, it usually turns out to be problematic to formalise.
This also makes the byte-size of the proof (about $24$kB) unfairly underestimated in comparison to the others.
\\
\textbf{The proof by Ho.}
In~\cite{ho2018word}, the first generalisation to the case $n>2$ was given.
The proof is remarkably short at around $11$kB. 
However, it crucially depends on a geometric, non-constructive result (\cite{burago1994periodic} and~\cite{alon1986borsuk}) not present in proof assistants' libraries.
Additionally, although more general than the previous and the present proofs, it pays its simplicity by using grammars of higher fanout than necessary, even in the known $n=2$ case.
This makes this approach not feasible for a formalisation.
\\
\textbf{The final proof in~\cite{gebhardt2022n}.}
This proof builds on that by~Ho and improves it by using grammars of the lowest possible fanout for all values of $n$.
The price to pay for that is a longer proof, of around $22$kB.
Additionally, this proof uses the Hobby-Rice theorem, not present in the \I{}/HOL library.
Interestingly, there is an extended version of the paper~\cite{gebhardt2020o_n} providing a Section~5 where the more geometric parts of the original proof are replaced by combinatorial equivalents, based on Tucker's and Ky Fan's lemmas.
Unfortunately, neither is not present in the \I{}/HOL library as well, and any of them would probably require a substantial amount of work.
This variation is also interesting because potentially impacting on criterion~\eqref{RefItemConstr}, since there are paper proofs of Tucker's lemma which are constructive.
For both variants, the nature of the dependencies and the size of the proofs would make their formalisation quite a complex task.

\section{Conclusions}
\label{RefSectConclusion}
A new proof for the fact that $O_2$ is a $2$-MCFL has been presented.
Contrary to the existing proofs, the present proof has no geometrical part whatsoever, and uses only elementary notions and methods.
This impacts both on its formalisability, being such proof comparably much more amenable to being formalised than existing ones, and on its computational interest, in that Lemma~\ref{RefLmBump} (expressing the informal argument of Section~\ref{RefSectIntui}) gives an extremely simple construction for a recursive implementation of a corresponding parser.
This paper therefore also serves as a starting point for a verified, efficient parser whose implementation is under way.
This is both practically and theoretically important not only from the computational linguistics point of view, but also from the point of view of computational algebra, where the word problem for $Z^2$ (tightly related to the parsing problem for $O_2$) has a prominent role.\cite{gilman2018groups}. 
While some of the existing proofs generalise beyond the $n=2$ case, that case is important because it is connected to formal languages theory, for example to the MIX language, and to combinatorial group theory problems such as the word problem~\cite{salvati2015mix}.
Therefore, this work, besides its proof-theoretical and algorithmic interest, provides a first step towards extending proof assistants' libraries in those two directions. 
Work is already underway to provide an executable formalisation in \I{}/HOL of the proposed proof, and a paper explaining the formalisation itself.
Lastly, it is hoped that a new proof and the novel concept involved, such as that of bump, can broaden the possible venues to tackle related open problems, such as the relationship between indexed languages~\cite{fischer1968PhD} and the word problem~\cite{gilman2018groups}.

\section*{Acknowledgments} 
The author would like to thank Mark Jan Nederhof for introducing the problem to him, and Matthew Barnes at Lancaster University Library for his meticulous and extremely helpful support in providing difficult to access resources. 
He would also like to thank Prof. Mark Brewer at Heilbronn and Prof. Constantin Blome for their academic mentorship. 
\newpage{}

\appendix{}
\section{Proofs}
\label{RefSectApp}
\begin{proof}[of Proposition~\ref{RefPropMain}]
The case $\left| pq \right| \leq 2$ is straightforward. Let us then consider a pair $\left( p, q \right)$ violating the thesis and such that $\left|  pq \right| \geq 4$ is the minimal possible.
\begin{enumerate}
\item{} Case $\left\{ p, q \right\} \cap O_2 = \emptyset$: 
\label{RefCaseBig}
By Theorem~\ref{RefLmMain}, let us consider a bal-decomposition of $\left( p, q \right)$ of components $s$ and $t$.
The length of either $s$ or $t$ must be $2$ since $p$ and $q$ are not balanced.
By symmetry of $G_2$, we can assume $\left| s \right|=2$, so that $s = \left( s_0, s_1 \right)$ and $I\left( s_0, s_1 \right)$ must be a sentential form by minimality.
If $\left| t \right| = 1$, so that $t =\left( t_0 \right)$, it must be $p = s_0 t_0$ and $q=s_1$, with $I \left( t_0, \epsilon \right)$ a sentential form, so that we could obtain that $I \left( s_0 t_0, s_1 \epsilon \right)$ is also a sentential form by, e.g., rule $ r_s $. 
Hence, we can assume $\left| s \right| = 2 = \left| t \right|$, and that $ I\left( s_0, s_1 \right)$, $ I \left( t_0, t_1 \right)$, $ I \left( s_0 s_1, \epsilon \right)$, $ I \left( t_0 t_1, \epsilon \right) $ are all sentential forms.
There are the following sub-cases, all leading to $ I \left( p, q \right)$ 
by applying the respective rule of $G_2$:
\begin{description}
\item [$ p = s_0 s_1 t_0$]: rule $ r_s $ to $I \left(s_0 s_1, \epsilon  \right)$ and $ I \left( t_0, t_1 \right)$. 
\item [$ p = s_0, q=s_1 t_0 t_1 $] : rule $ r_s $ to $I \left(s_0, s_1  \right)$ and $ I \left( \epsilon, t_0 t_1 \right)$.
\item [$ p = s_0, q=t_0 s_1 t_1$]: rule $r_r$  to $I \left(s_0, s_1  \right)$ and $ I \left( t_0, t_1 \right)$.
\item [$ p = s_0 t_0 s_1$]: rule $r_l$  to $I \left(s_0, s_1  \right)$ and $ I \left( t_0, t_1 \right)$.
\item [$ p = s_0 t_0, q= t_1 s_1$]: rule $r_n$  to $I \left(s_0, s_1  \right)$ and $ I \left( t_0, t_1 \right)$.
\item [$ p = s_0 t_0, q= s_1 t_1$]: rule $r_s$  to $I \left(s_0, s_1  \right)$ and $ I \left( t_0, t_1 \right)$.
\end{description}
\item{} Case $\left\{ p, q \right\} \subseteq O_2 \sdiff \left\{ \epsilon \right\}$: then $I \left( p, \epsilon \right)$ and $ I \left( \epsilon, q\right)$ are both sentential forms by minimality, so that we can just apply rule $ r_n $ to obtain that $ I \left( p, q \right)$ is a sentential form.
\item{} Case $\epsilon \in \left\{ p, q \right\}$: if $q= \epsilon$, then $ I \left( p \left( 0 \right), \factor{p}{1}{} \right)$ is a sentential form by case~\eqref{RefCaseBig}, as is $ I \left( \epsilon, \epsilon \right)$ using $r_0$.
Hence $ I\left( p \left( 0 \right) \epsilon \factor{p}{1}{}, \epsilon \right)$ is also a sentential form by rule $r_l$. 
Similarly if $p= \epsilon$. 
\end{enumerate}
\end{proof}

\begin{proof}[of Corollary~\ref{RefLmG2O2}]
It suffices to show that $O_2 \subseteq L \left( G_2 \right)$. Any non-empty string $s$ of $O_2$ is in $L \left( G_2 \right)$ since that $ I \left( s_0, \factor{s}{1}{} \right)$ is a sentential form by~Proposition~\ref{RefPropMain} and by applying rule~$r_z$ to $ I \left( s_0, \factor{s}{1}{} \right)$.
\end{proof}


\bibliographystyle{splncs04}
\bibliography{mbc}
\end{document}